\documentclass[pra,twocolumn,eqsecnum,superscriptaddress,showpacs,amsmath,amstex,amssymb,citeautoscript]{revtex4-1}
\usepackage{bm}

\usepackage{amsmath,amssymb,bm,bbm} 
\usepackage{graphicx}  
\usepackage{color} 
\usepackage[dvipsnames]{xcolor}
\usepackage[papersize={8.5in,11in}]{geometry}
\usepackage[colorlinks=true]{hyperref}  
\usepackage{ulem}
\usepackage[mathscr]{euscript}
\usepackage[section]{placeins}
\hypersetup{
    bookmarks=true,         
    unicode=false,          
    pdftoolbar=true,        
    pdfmenubar=true,        
    pdffitwindow=false,     
    pdfstartview={FitH},    
    pdfsubject={},   
    pdfcreator={},   
    pdfproducer={}, 
    pdfkeywords={} {} {}, 
    pdfnewwindow=true,      
    colorlinks=true,       
    linkcolor=magenta, 
    citecolor=blue,        
    filecolor=magenta,      
    urlcolor=blue           
} 

\usepackage{amsfonts}
\usepackage{upgreek}
\usepackage{slashed}
\usepackage{latexsym}

\newcommand{\beq}{\begin{equation}}
\newcommand{\eeq}{\end{equation}}
\def\bea{\begin{eqnarray}}
\def\eea{\end{eqnarray}}

\usepackage{multirow}

\usepackage{braket}
\usepackage{comment}
\usepackage{slashed}
\usepackage{subcaption}

\usepackage{comment}
\usepackage{pifont}

\usepackage{relsize}

\usepackage{soul}

\usepackage{pgfplots}
\usepackage{ifthen}

\begin{document}

\author{Harley Scammell}
\affiliation{School of Mathematical and Physical Sciences, University of Technology Sydney, Ultimo, NSW 2007, Australia}

\author{Oleg P. Sushkov}
\affiliation{School of Physics, University of New South Wales, Kensington, NSW 2052, Australia}

\date{\today}

\title{Zener tunnelling in biased bilayer graphene via analytic continuation of\\ semiclassical theory}

\begin{abstract}
Employing a semiclassical method based on analytic continuation, we compute the electron–hole pair production rate in biased bilayer graphene subject to an in-plane electric field. This approach, originally due to Zwaan, bypasses the need for exact solutions at turning points, which are generally unavailable beyond linear or quadratic band structures. Applying this technique to biased bilayer graphene reveals non-standard features of the asymptotic wavefunctions, in particular the necessity of retaining decaying components even in classically allowed regions. By providing a fully analytic solution, this work complements and clarifies earlier results based on hybrid analytical–numerical treatments, and importantly establishes the absolute normalisation of the pair production rate—and hence of the tunnelling current.
\end{abstract}

\maketitle

\section{Introduction}

Bilayer graphene subject to a perpendicular electric field—i.e., biased bilayer graphene (Bblg)—develops a tunable gap in its electronic spectrum, becoming a semiconductor at charge neutrality~\cite{Zhang2009}, and with dual gating the carrier density can be independently controlled~\cite{Taychatanapat2010,engdahl2025}. This electric field tunability has made Bblg an attractive platform for predicting and observing various many-body quantum phenomena, including momentum-polarised condensation~\cite{Jung2015,LinLi2023,Dong2023}, exciton condensation~\cite{Scammell_EC}, and a cascade of isospin magnetic orders~\cite{delaBarrera2022,Zhou2022}. 
At the single-particle level, strong in-plane electric fields can also drive interband Zener tunnelling across the gap—an effect that is both theoretically rich and technologically relevant.

Zener tunnelling refers to the non-perturbative process in which electrons tunnel from the valence to the conduction band under the influence of a strong electric field~\cite{Zener}. More broadly, tunnelling induced by strong electric and magnetic fields has long been studied in quantum electrodynamics, beginning with early work by Sauter~\cite{Sauter1931} and formalised in the Heisenberg–Euler effective action~\cite{HeisenbergEuler1936}. In low-dimensional systems such as bilayer graphene, it is both experimentally accessible and theoretically rich, offering insight into quantum interference, nonlinear response, and field-induced transport regimes~\cite{LevitovPNAS2011,OkaAoki2005,KatsnelsonBLG2012}. In Bblg specifically, the presence of a gap makes Zener tunnelling particularly relevant to prospective device applications. Tunnel field-effect transistors (TFETs) based on interband tunnelling are among the leading candidates for steep-slope, low-power switching~\cite{Ghosh2013,QinSteep}.


The specific question we address in this work is: \textit{Given an in-plane electric field, what is the pair production rate across the gap of Bblg?} This problem was previously studied by Nandkishore and Levitov~\cite{LevitovPNAS2011}, who employed a combination of the WKB approximation and numerical solutions to the relevant Schr\"odinger equation. In their approach, unknown amplitudes and scattering phases in the WKB framework were extracted by fitting to the numerical solution.



In this work we present an alternative and fully analytic method, based on semiclassical theory extended by analytic continuation—originally introduced by Zwaan~\cite{Zwaan1942}.  
Although developed in the context of atomic physics, this approach has remained largely unexplored in condensed matter settings. It bypasses the need for local exact solutions near turning points (e.g. Airy functions in the quadratic case) and instead constructs globally valid solutions via continuation through the complex plane. This is particularly valuable for systems like Bblg, where the non-quadratic dispersion precludes known closed-form solutions near turning points.

The key motivation and advances of this paper are as follows. First, the method is conceptually distinct from the WKB approach of Ref.~\cite{LevitovPNAS2011} and provides a route for computing both the scattering phase and tunnelling coefficient within a single, analytic framework. Second, its application to Bblg reveals subtle but important features of the wavefunction structure—notably, the necessity of retaining exponentially decaying components in classically {\it allowed} regions. 
Third, this method allows the definition of the absolute normalisation of the pair-production rate, and with it, the correct expression for the tunnelling current.
Relatedly, we highlight subtleties related to the normalisation of the Dirac sea. 


This paper is structured to equip the interested reader with all necessary details. Section~\ref{sec:outline} outlines the key steps of the method. Section~\ref{sec:results} presents the full derivation of the semiclassical wavefunctions and matching conditions. Section~\ref{sec:decay} details the proper normalisation of states and its dependence on device geometry, culminating in an expression for the pair production rate in absolute units. Section~\ref{sec:discussion} explores the physical implications of these results. Finally, Appendix~\ref{sec:Dirac} provides an analogous treatment for Zener tunnelling in the gapped two dimensional (2D) Dirac Hamiltonian.

\section{Outline of the method}\label{sec:outline}

\textbf{Standard structure of solutions:} In conventional Zener tunnelling problems with quadratic band edges, the semiclassical wavefunction exhibits the following structure:
\begin{itemize}
\item Oscillatory (running wave) solutions in the classically allowed region $x > x_+$, which carry the tunnelling current.
\item Exponentially decaying solutions in the classically forbidden region $x_- < x < x_+$.
\item Standing waves (a superposition of left- and right-moving components) in the classically allowed region $x < x_-$.
\end{itemize}

\textbf{Analytic continuation approach:} The semiclassical approximation breaks down near the turning points at $x = x_\pm$, where the classical momentum vanishes. Zwaan's key insight is to avoid these singular points by staying a small distance away—at $x - x_\pm = \rho$, with $\rho$ finite—and to match the solutions by analytically continuing around the turning points in the complex plane.

The continuation is performed by rotating the variable as $\rho \to \rho e^{i\phi}$ while keeping $|x - x_\pm| = \rho$ fixed. This allows one to bridge the solutions on either side of the turning point (i.e., $x - x_\pm > 0$ and $x - x_\pm < 0$) without leaving the domain of semiclassical validity. Both clockwise and counterclockwise continuation paths must be considered to enforce the cancellation of unphysical, exponentially growing components. The same procedure is carried out near both $x = x_+$ and $x = x_-$.

\begin{figure}[t!]
\includegraphics[width=0.35\textwidth]{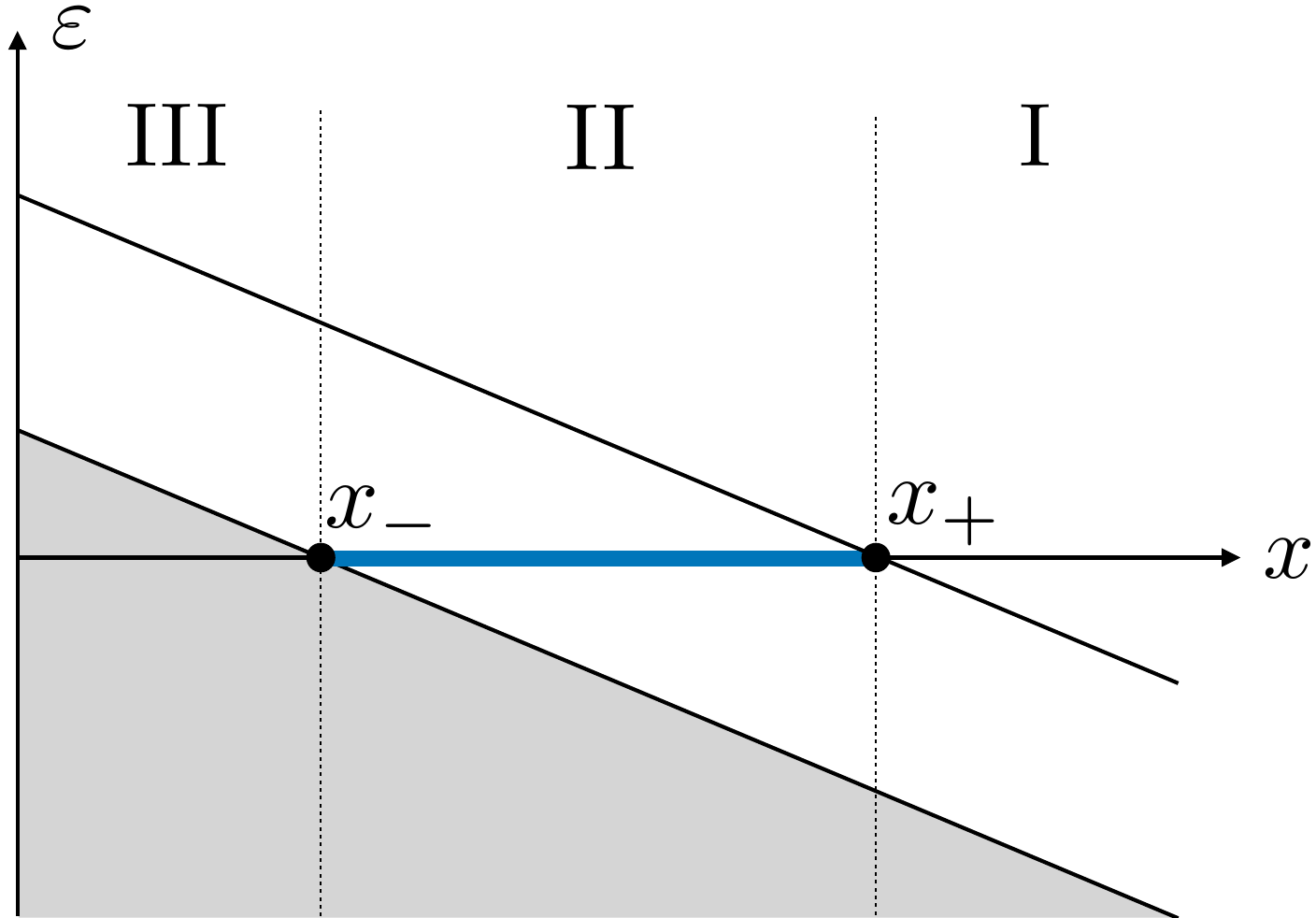}
\caption{Semiclassical energy spectrum in Bblg, showing classically allowed and forbidden regions. The grey shaded area represents the filled valence band; the black diagonal line indicates the conduction band. The horizontal axis corresponds to the spatial coordinate $x$ at fixed energy $\varepsilon=0$. The turning points $x_\pm$ mark the boundaries between regions I, II, and III.}
\label{fig:regions}
\end{figure}

\textbf{Modified structure for Bblg:} For biased bilayer graphene, the semiclassical wavefunction exhibits a richer structure due to its non-quadratic dispersion. As we will show:
\begin{itemize}
\item In region I ($x > x_+$), both running and decaying solutions appear.
\item In region II ($x_- < x < x_+$), two linearly independent decaying solutions exist.
\item In region III ($x < x_-$), both standing and decaying wavefunctions are present.
\end{itemize}

By analytically continuing the semiclassical wavefunction around each turning point and carefully matching across the boundaries, we obtain a globally consistent solution across all regions. The tunnelling probability then follows from the asymptotic amplitudes of the running wave in region I.

\section{Wavefunction Matching via Analytic Continuation}\label{sec:results}
\subsection{Semiclassical solutions in different regions}
Schr\"odinger equation for Bblg, at a given valley, subject to in-plane electric field $F$, reads
\begin{eqnarray}
  \label{SE}
  \left[
    \left(
    \begin{array}{cc}
      \Delta & \frac{p_+^2}{2m}\\
      \frac{p_-^2}{2m} &-\Delta
      \end{array}
      \right) -Fx\right]\psi=\varepsilon\psi.
\end{eqnarray}
The filling of states is shown in Fig. \ref{fig:regions}. 
Writing the wave function as
\begin{align}
\label{wavefunc}
\psi(x,y)=e^{ik_y y}\begin{pmatrix} a(x) \\ b(x)\end{pmatrix},
\end{align}
 the $x$-component of the probability current is
\begin{eqnarray}
\label{jx1}  
  j_x&=\frac{1}{2m}\text{Im}\big[(\partial_xa^*(x))b(x)+(\partial_xb^*(x))a(x)\big].
\end{eqnarray}  
The semiclassical $x$-momentum equation is
\begin{eqnarray}
  \label{psx}
  p(x)^2&=&\pm 2m\sqrt{F^2x^2-\Delta^2}-k_y^2.
\end{eqnarray}
Here we set the energy $\varepsilon=0$ since it is equivalent to the shift of $x$.
The solution with sign ($-$) implies that the momentum is complex.
This is true even in the classically allowed regions, $x \to \pm\infty$.
In our analytic treatment, we are able to restrict the Hilbert space to {\it physical} wave functions, i.e. not exponentially growing. We therefore throw away exponentially growing solutions coming from sign ($-$). We note that the existence of such solution leads to problems
with a direct numerical treatment. The physical solution with sign ($+$) implies that the momentum
is real at $x \to \pm\infty$ and there are two turning points
\begin{eqnarray}
\label{tp}  
x_{\pm}=\pm \frac{1}{F}\sqrt{\Delta^2+\frac{k_y^4}{4m^2}}.
\end{eqnarray}

\subsubsection{Notation and definitions}
Before proceeding, we define the classical momentum in the forbidden and allowed regions, denoted $p_F(x)$ and $p_A(x)$ respectively, as
\begin{align}
\label{p_FA}
\notag p_F(x)&\equiv\sqrt{2m}(\Delta^2-F^2x^2)^{1/4},\\ 
p_A(x)&\equiv \sqrt{2m}(F^2x^2-\Delta^2)^{1/4}.
\end{align}
In the wave function expressions, we will also make repeated use of 
\begin{align}
\label{alpha_beta}
\alpha_A(x)&\equiv \left(\frac{Fx+\Delta}{Fx-\Delta}\right)^{1/4}, \quad \alpha_F(x)\equiv \left(\frac{\Delta+Fx}{\Delta-Fx}\right)^{1/4}
\end{align}
which are clearly related by a complex phase exp$\{i\pi/4\}$, yet we find it convenient to distinguish since they are each real functions in the allowed and forbidden regions, respectively.

Finally, we make repeated use of
\begin{eqnarray}
  \label{tun2}
 S_0&=&\int_{x_-}^{x_+}p_F(x)dx=\beta_0\sqrt{2m}\frac{\Delta^{3/2}}{F}
 \nonumber\\
 S_y&=&\int_{x_-}^{x_+}\frac{k_y^2}{2p_F(x)}dx=
 \beta_1\frac{k_y^2}{F}\sqrt{\frac{\Delta}{2m}}\nonumber\\
 \beta_0&=&1.748\ , \ \ \ \beta_1=1.198.
 \end{eqnarray}  
For $ S_y$ we expand up to the leading order in $k_y^2$. For clarity, we will perform the majority of calculations at $k_y=0$, yet to recover non-zero $k_y$, the following rule is required
\begin{align}
\label{rule}
e^{\pm i\pi/4}[S_0]\to e^{\pm i\pi/4}[S_0\mp i S_y].
\end{align}

\subsubsection{Region I: $x > \Delta/F$}
Let us consider for now the case $k_y=0$.
Semiclassical solution for the running escaping wave at $x > \Delta/F$,  i.e. a classically allowed region, is
\begin{eqnarray}
  \label{rw1}
  a(x)&=&\frac{T}{2\sqrt{p_A(x)}}\alpha_A(x)e^{iS}\nonumber\\
  b(x)&=&\frac{T}{2\sqrt{p_A(x)}}\alpha_A(-x)e^{iS}\nonumber\\
  p_A(x)&=&\sqrt{2m}(F^2x^2-\Delta^2)^{1/4}\nonumber\\
   S&=&\int_{\Delta/F}^{x}p_A(x^\prime) dx^\prime.
\end{eqnarray}
Here $T$ is the transmission amplitude.
The running wave carries current $j_x=|T|^2/(2m)$.
The semiclassical approximation is justified when the distance from the
turning point is march larger than the wavelength $1/p_A(x)$, so one can say that
this is the large mass approximation. The semiclassical expansion is the
expansion in powers of $1/p_A(x)$.

Exponentially decaying solution  at $x > \Delta/F$ reads 
\begin{eqnarray}
  \label{ed1}
  a(x)&=&\frac{1}{\sqrt{2p}}\alpha_A(x)e^{-S}\nonumber\\
  b(x)&=&\frac{-1}{\sqrt{2p}}\alpha_A(-x)e^{-S}\nonumber\\
   S&=&\int_{\Delta/F}^{x}p_A(x^\prime) dx^\prime.
\end{eqnarray}
The normalisation of the decaying solution is arbitrary at this point in the analysis; later it will be fixed by matching conditions at the turning points. 

\subsubsection{Region II:  $-\Delta/F < x < \Delta/F$}
In the classically forbidden region, $-\Delta/F < x < \Delta/F$ there are two
exponentially decaying solutions
\begin{eqnarray}
  \label{ew1}
  a(x)&=&\frac{1}{\sqrt{2p_F(x)}}\alpha_F(x)\exp(-e^{\pm i\pi/4}S)\nonumber\\
  b(x)&=&\frac{\mp i}{\sqrt{2p_F(x)}}\alpha_F(-x)\exp(-e^{\pm i\pi/4}S)\nonumber\\
   S&=&\int_{-\Delta/F}^{x}p_F(x^\prime) dx^\prime.
\end{eqnarray}
The sign $\pm$ in the phase correlates with the sign $\mp$ in $b(x)$.
Of course the solutions (\ref{ew1}) can be multiplied by arbitrary constants.
Note that while the solutions (\ref{ew1})  have  complex $x$-dependent
action $-e^{\pm i\pi/4}S$ nevertheless each solution carries zero current, $j_x=0$.

\subsubsection{Region III:  $ x < -\Delta/F$}
In the classically allowed region, $x< -\Delta/F$, there is the following
standing wave solution
\begin{eqnarray}
  \label{sw1}
  a(x)&=&\frac{1}{\sqrt{p}_A(x)}\alpha_A(x)\cos(S+\phi)\nonumber\\
  b(x)&=&-\frac{1}{\sqrt{p}_A(x)}\alpha_A(-x)\cos(S+\phi)\nonumber\\
   S&=&\int_{x}^{-\Delta/F}p_A(x^\prime) dx^\prime.
\end{eqnarray}
Note that the standing wave is normalised is such a way that at $x\to -\infty$
the averaged over oscillations probability is
\begin{eqnarray}
  \label{norm}
  {\overline {a(x)^2+b(x)^2}}\to\frac{1}{p_A(x)}\to\frac{1}{\sqrt{2mF|x|}}.
\end{eqnarray}

The exponentially decaying solution in the classically allowed region, $x< -\Delta/F$, reads 
\begin{eqnarray}
  \label{edl}
  a(x)&=&\frac{1}{\sqrt{2p}}\alpha_A(x)e^{-S}\nonumber\\
  b(x)&=&\frac{1}{\sqrt{2p}} \alpha_A(-x)e^{-S}\nonumber\\
     S&=&\int_{x}^{-\Delta/F}p_A(x^\prime) dx^\prime.
\end{eqnarray}

\subsection{Analytic continuation through turning points in complex $x$-plane.}
\begin{figure}[t!]
\includegraphics[width=0.45\textwidth]{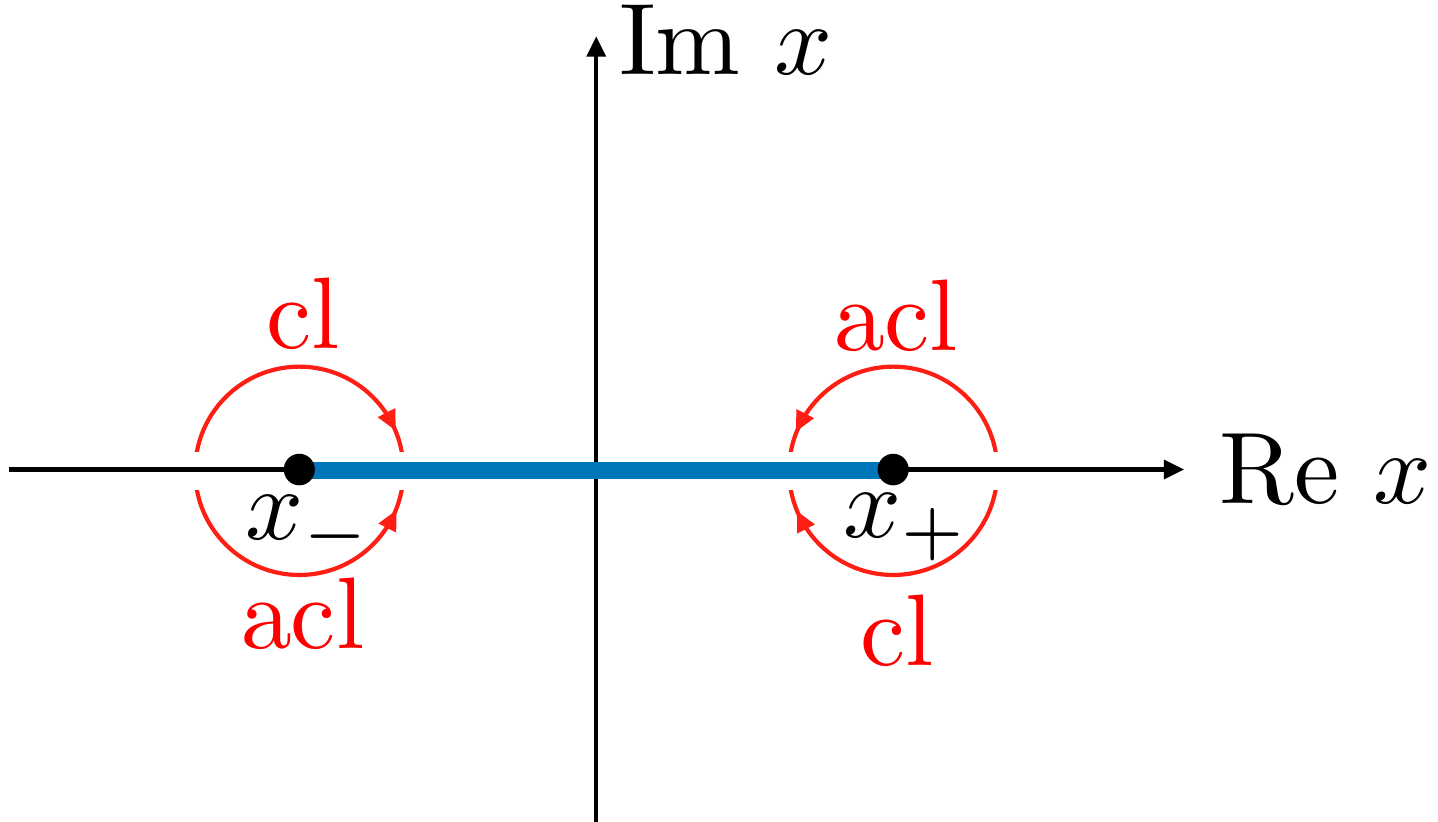}
\caption{Analytic continuation through turning points in complex $x$-plane. The two distinct path directions at each turning point are indicated by the arrowed red lines and labels cl (clockwise) and acl (anti-clockwise).}
\label{cut}
\end{figure}

We employ the standard method of analytical continuation of semiclassical
solutions in complex $x$-plane\cite{LL3}.
There are four ways shown in Fig.~\ref{cut} for continuation  from the
classically allowed regimes to the classically forbidden one.\\

\noindent
{\bf Analytic continuation through $x_+$:} Let us first perform the analytic continuation of the running wave
(\ref{rw1}) from $x > x_+$ to  $x < x_+$.  Doing so we perform the
rotation $x-x_+=\rho e^{i\phi}$. The anticlockwise (acl), $\phi : 0 \to  \pi$, and the clockwise (cl) ,  $\phi : 0 \to -\pi$, rotations are possible.
The action $\int p_A(x) dx \propto \rho^{5/4}e^{i5\phi/4}$, hence
the exponent in (\ref{rw1}) is changed as
\begin{eqnarray}
  \label{ac1}
  &&\text{acl}: \ \ \ iS \to ie^{i5\pi/4}S=e^{-i\pi/4}S  \nonumber\\
  &&\text{cl}: \ \ \ iS \to ie^{-i5\pi/4}S=-e^{i\pi/4}S.
  \end{eqnarray}
In our conventions, $S$ is always positively defined.
Therefore the ``cl'' continuation decays exponentially inside the classically
forbidden region and hence has to be thrown away.
The action S in (\ref{ac1}) is taken with respect to the point $x_+$.
One can shift the reference point to $x_-$ by replacing $S\to S_0-S$.
Hence the exponential factor in the ``acl'' analytically continued solution
is 
\begin{eqnarray}
  \label{efact}
  exp\left\{e^{-i\pi/4}S_0\right\} exp\left\{-e^{-i\pi/4}S\right\}
\end{eqnarray}
where S is calculated with respect to $x_-$.
The prefactors in (\ref{rw1}) continue as
\begin{align}
  \label{ac2}
  \text{acl}&: \ \  a\propto (x-x_+)^{-1/4}(x-x_+)^{-1/8}\to e^{-i3\pi/8}\nonumber\\
  \text{acl}&: \ \  b\propto (x-x_+)^{+1/4}(x-x_+)^{-1/8}\to e^{+i\pi/8}.
\end{align}
Hence, the analytic continuation of the running wave at $x> x_+$ leads to the
following wave at $x< x_+$
\begin{eqnarray}
  \label{ac3}
  \varphi_{acl}^{(r)}&=&{\cal P}_T\frac{e^{\left\{-e^{-i\frac{\pi}{4}}S\right\}}}{\sqrt{2p_F(x)}}
\left(
  \begin{array}{c}
\alpha_A(x)\\
 i\alpha_A(-x)
  \end{array}
  \right)
  \nonumber\\
        {\cal P}_T&=&\frac{T}{\sqrt{2}}e^{-i\frac{3\pi}{8}}e^{\left\{e^{-i\frac{\pi}{4}}S_0\right\}}
        \nonumber\\
        S&=&\int_{-\Delta/F}^{x}p_F(x^\prime) dx^\prime.
  \end{eqnarray}
Up to the prefactor ${\cal P}$ this is one of the solutions (\ref{ew1}).\\

Similar continuation of the decaying wave (\ref{ed1}) gives two solutions
in the classically forbidden region
\begin{eqnarray}
  \label{ac3d}
  \varphi_{acl}^{(d)}&=&{\cal P}_{acl}\frac{e^{\left\{-e^{i\frac{\pi}{4}}S\right\}}}{\sqrt{2p_F(x)}}
\left(
  \begin{array}{c}
\alpha_F(x)\\
 -i\alpha_F(-x)
  \end{array}
  \right)\nonumber\\
{\cal P}_{acl}&=&e^{-i\frac{3\pi}{8}}e^{\left\{e^{i\frac{\pi}{4}}S_0\right\}}\nonumber\\
  \varphi_{cl}^{(d)}&=&{\cal P}_{cl}\frac{e^{\left\{-e^{-i\frac{\pi}{4}}S\right\}}}{\sqrt{2p_F(x)}}
\left(
  \begin{array}{c}
\alpha_F(x)\\
 i\alpha_F(-x)
  \end{array}
  \right)\nonumber\\
{\cal P}_{cl}&=&e^{i\frac{3\pi}{8}}e^{\left\{e^{-i\frac{\pi}{4}}S_0\right\}}.
\end{eqnarray}
Here $S$ and $p_F(x)$ coincide with that in (\ref{ac3}).
Of course up to coefficients these solutions coincide with (\ref{ew1}).\\

\noindent
{\bf Analytic continuation through $x_-$:}
The standing wave (\ref{sw1}) has two components $e^{+iS}$ and $e^{-iS}$
The rotation is , see Fig.~\ref{cut}, $x_--x=\rho e^{i\phi}$, and the action
rotates as  $S \propto \rho^{5/4}e^{i5\phi/4}$.
Therefore under the rotations the action changes as
\begin{eqnarray}
  \label{rl}
  &&acl: \ \ \ iS \to ie^{i5\pi/4}S=e^{-i\pi/4}S  \nonumber\\
  &&cl: \ \ \ iS \to ie^{-i5\pi/4}S=-e^{i\pi/4}S.
  \end{eqnarray}
Hence, the $e^{+iS}$ part is transferred to
\begin{eqnarray}
  \label{Sp}
  && e^{+iS} \ \ \ part:\\
  &&acl: \ \  \frac{e^{i\pi/8+i\phi}}{\sqrt{2}} \frac{\exp\left\{e^{-i\pi/4}S\right\}}{\sqrt{2p_F(x)}} 
  \left(
  \begin{array}{c}
\alpha_F(x) \\  
i\alpha_F(-x)
  \end{array}
  \right),\nonumber\\
    &&cl: \ \  \frac{e^{-i\pi/8+i\phi}}{\sqrt{2}} \frac{\exp\left\{-e^{i\pi/4}S\right\}}{\sqrt{2p_F(x)}} 
  \left(
  \begin{array}{c}
\alpha_F(x) \\  
-i\alpha_F(-x)
  \end{array}
  \right).\nonumber 
\end{eqnarray}
The $e^{-iS}$ part is transferred to
\begin{eqnarray}
  \label{Sm}
  && e^{-iS} \ \ \ part:\\
    &&acl: \ \  \frac{e^{i\pi/8-i\phi}}{\sqrt{2}}
    \frac{\exp\left\{-e^{-i\pi/4}S\right\}}{\sqrt{2p_F(x)}} 
  \left(
  \begin{array}{c}
\alpha_F(x) \\  
i\alpha_F(-x)
  \end{array}
  \right),\nonumber\\
    &&cl: \ \  \frac{e^{-i\pi/8-i\phi}}{\sqrt{2}}
    \frac{\exp\left\{e^{i\pi/4}S\right\}}{\sqrt{2p_F(x)}} 
  \left(
  \begin{array}{c}
\alpha_F(x) \\  
-i\alpha_F(-x)
  \end{array}
  \right).\nonumber
\end{eqnarray}
Similar continuation of the decaying wave $e^{-S}$, Eq.~(\ref{edl}) gives
the following  solutions in the classically forbidden region
\begin{eqnarray}
  \label{Sd}
  && e^{-S} \ \ \ part:\\
    &&acl: \ \  e^{i\pi/8}
    \frac{\exp\left\{e^{i\pi/4}S\right\}}{\sqrt{2p_F(x)}} 
  \left(
  \begin{array}{c}
\alpha_A(x) \\  
-i\alpha_A(-x)
  \end{array}
  \right),\nonumber\\
    &&cl: \ \  e^{-i\pi/8}
    \frac{\exp\left\{e^{-i\pi/4}S\right\}}{\sqrt{2p_F(x)}} 
  \left(
  \begin{array}{c}
\alpha_A(x) \\  
i\alpha_A(-x)
  \end{array}
  \right).\nonumber
\end{eqnarray}

\subsection{Matching solutions in the classically forbidden region.
Transmission amplitude}

\noindent
    {\bf Matching at $x_-$:}
    The wave function (\ref{Sd}) grows exponentially at $ x > x_-$, at the
    same  time the wave functions (\ref{Sp}) and (\ref{Sm}) also contain
      exponentially growing components. 
      The correct wave function is a linear combination
      \begin{eqnarray}
  \label{lcl}      
\Psi_L=|\cos(S)\rangle+A|e^{-S}\rangle        
\end{eqnarray}
where the exponentially growing components are cancelled out.
This condition results in two equations
\begin{eqnarray}
  \label{cond1}
&&  \frac{1}{\sqrt{2}}e^{i\pi/8+i\phi}+Ae^{-i\pi/8}=0,\nonumber\\
  &&  \frac{1}{\sqrt{2}}e^{-i\pi/8-i\phi}+Ae^{i\pi/8}=0.
\end{eqnarray}
Therefore, the scattering phase $\phi$ and the coefficient A are
\begin{eqnarray}
  \label{scat}
  \phi=-\frac{\pi}{4}\ , \ \ \ A=-\frac{1}{\sqrt{2}}.
\end{eqnarray}
Hence, using (\ref{lcl}), (\ref{Sp}), and (\ref{Sm}) we find the
{\it left} wave function, $\Psi_L$, under the barrier
\begin{align}
  \label{lcl1}      
  \Psi_L&=
e^{-i3\pi/8} \frac{\exp\left\{-e^{i\pi/4}S\right\}}{2\sqrt{p_F(x)}} 
  \left(
  \begin{array}{c}
\alpha_F(x) \\  
-i\alpha_F(-x)
  \end{array}
  \right)\nonumber\\
  & \ \ +e^{i3\pi/8}
    \frac{\exp\left\{-e^{-i\pi/4}S\right\}}{2\sqrt{p_F(x)}} 
  \left(
  \begin{array}{c}
\alpha_F(x) \\  
i\alpha_F(-x)
  \end{array}
  \right).
\end{align}

\noindent
    {\bf Matching at $x_+$:}
    The wave function coming from the right is a linear combination of the
running wave (\ref{rw1}) and the exponentially decaying wave (\ref{ed1})
      \begin{eqnarray}
  \label{lcr}      
\Psi_R=|e^{iS}\rangle+B|e^{-S}\rangle        
\end{eqnarray}
where $B$ is a coefficient.
Analytic continuations of $|e^{iS}\rangle$ and $|e^{-S}\rangle$ to the sub-barrier
region are given by (\ref{ac3}) and (\ref{ac3d}).
Hence
 \begin{align}
\label{lcr1}
&\hspace{-0.13cm}\Psi_R = 
\frac{e^{e^{+i\frac{\pi}{4}}S_0}}{\sqrt{2p_F(x)}}\, 
Be^{-i\frac{3\pi}{8}}\, e^{-e^{i\frac{\pi}{4}}S}
\begin{pmatrix}
\alpha_F(x)\\
 -i\alpha_F(-x)
\end{pmatrix}\\
&+
\frac{e^{e^{-i\frac{\pi}{4}}S_0}}{\sqrt{2p_F(x)}}\,
\left[\frac{T}{\sqrt{2}}e^{-i\frac{3\pi}{8}}+Be^{i\frac{3\pi}{8}}\right]
e^{-e^{-i\frac{\pi}{4}}S}
\begin{pmatrix}
\alpha_F(x)\\
 i\alpha_F(-x)
\end{pmatrix}\nonumber
\end{align}
 which must coincide with (\ref{lcl1}). We thereby solve the system of equations
 \begin{eqnarray}
   \label{leq}
   &&\frac{1}{\sqrt{2}}=e^{\left\{e^{-i\frac{\pi}{4}}S_0\right\}}
   \left[\frac{e^{-i3\pi/4}}{\sqrt{2}}T+B\right],\nonumber\\
   &&\frac{1}{\sqrt{2}}=e^{\left\{e^{i\frac{\pi}{4}}S_0\right\}}B,
   \end{eqnarray}
to find the transmission amplitude
 \begin{eqnarray}
   \label{tr}
   T=-2e^{i\pi/4}e^{-S_0/\sqrt{2}}\sin(S_0/\sqrt{2}).
 \end{eqnarray}
 Here $S_0$ is given in Eq.~(\ref{tun2}).

Eq. (\ref{tr}) has been derived for $k_y=0$. According to the rule \eqref{rule}, with account of $k_y$ the exponent $e^{i\pi/4}S_0$ in
Eq.~(\ref{leq}) has to be replaced by $e^{i\pi/4}(S_0-iS_y)$ and 
the exponent $e^{-i\pi/4}S_0$ in has to be replaced by $e^{i\pi/4}(S_0+iS_y)$.
After that solution of (\ref{leq}) gives the tunnelling amplitude at a nonzero
$k_y$
 \begin{align}
   \label{trk}
   T(k_y)=-2e^{i\pi/4}e^{-(S_0+S_y)/\sqrt{2}}\sin\left(\frac{S_0-S_y}{\sqrt{2}}\right).
 \end{align}

 \section{Pair production rate}\label{sec:decay}

\begin{figure}
    \centering
\includegraphics[width=0.35\textwidth]{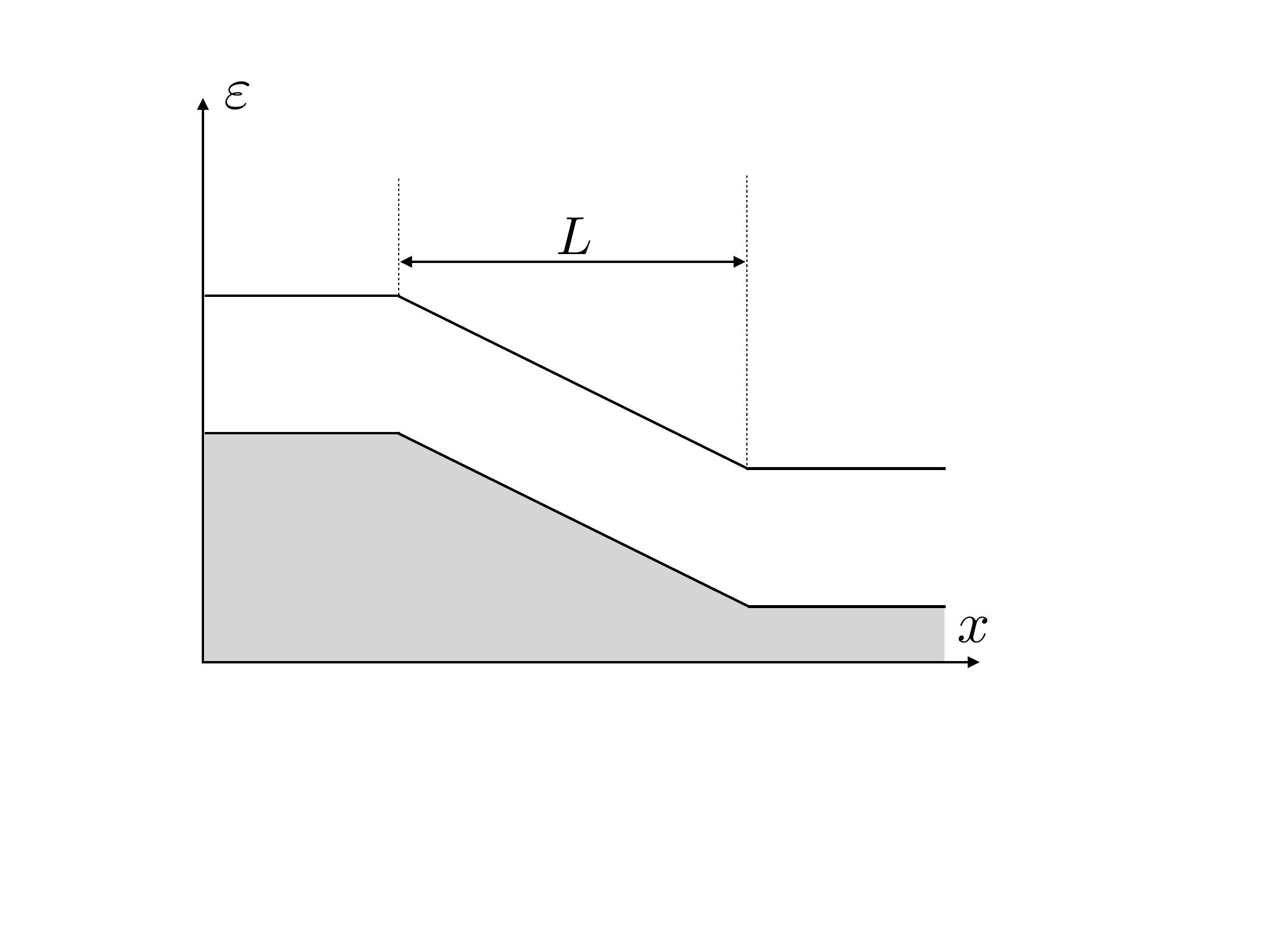}
    \caption{Device setup: application of field $F$ is restricted to a finite length, $L$.}
    \label{fig:device}
\end{figure}

Another peculiarity of bilayer graphene is that the normalisation of wave functions depends on the sample length $L$ along the direction of the electric field. In this subsection, we detail the correct normalisation of standing wave states, and the subsequent pair production rate, for a device of length $L$,  as depicted in Fig. \ref{fig:device}. We will consider two regimes, {\it long} and {\it short devices}, which we make precise now. 

As discussed in Ref.~\cite{McCann2013}, for $\Delta = 0$, the low-energy dispersion exhibits a crossover at momentum scale $p \approx \gamma_1 / (2v)$ from quadratic to linear behaviour. Let us denote the corresponding energy scale $\Delta_1 = \gamma_1 /2 \approx 380$ meV. At small momentum, the spectrum is approximately quadratic, $\varepsilon \approx p^2 / 2m$, while at large momentum it becomes linear, $\varepsilon \approx v p$. The velocity and effective mass are given by $v = \sqrt{3} a \gamma_0 /2$
and $m = \gamma_1 / 2v^2$, respectively, where $\gamma_0$ and $\gamma_1$ are Slonczewski–Weiss–McClure model parameters~\cite{McCann2013} and we work in units with $\hbar=1$.

\begin{figure}[t!]
  \includegraphics[width=0.45\textwidth]{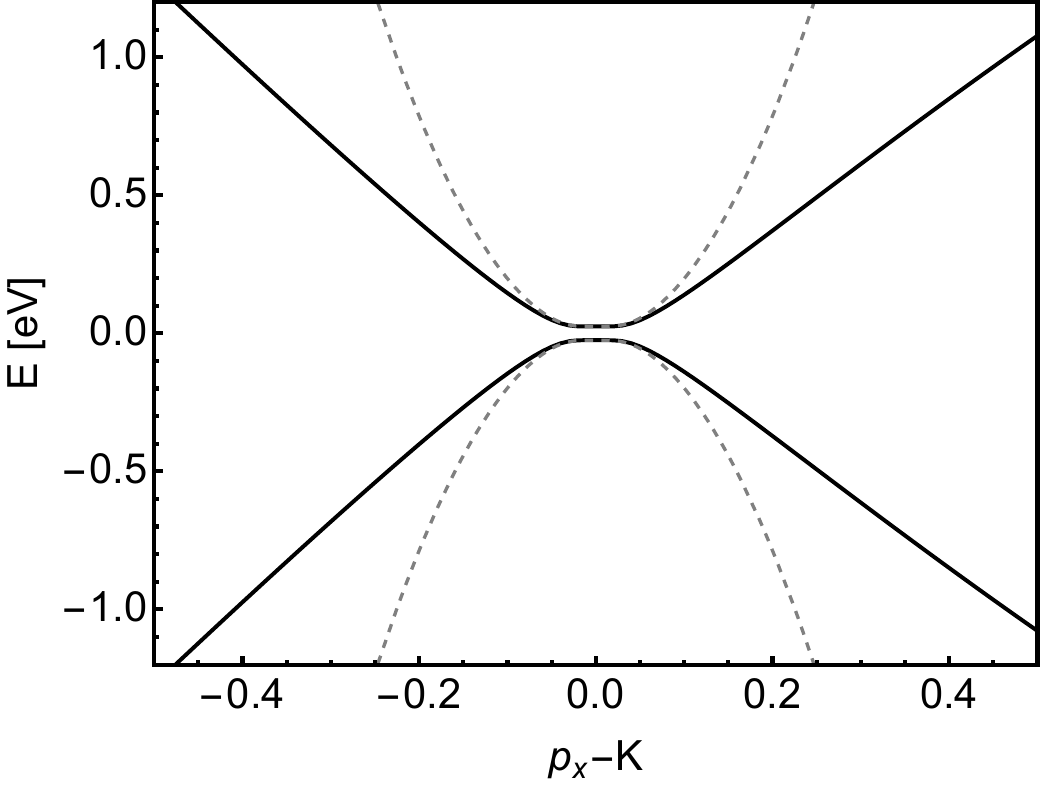}
  \caption{Two branches of the dispersion
near the K-point, $K=\frac{4\pi}{3a}$. The band gap $2\Delta=50meV$. 
Dashed lines show the dispersion of the approximation $\pm \sqrt{p^4/(4m^2)+\Delta^2}$.
}    
\label{disp}
\end{figure}

To observe this crossover, we present in Fig.~\ref{disp} the band dispersion of the tight binding model \cite{McCann2013} as well as the bands of \eqref{SE} at $F=0$.

\subsubsection{Normalisation in the linear regime (long device)}
Consider the sufficiently long sample, $L$, such that we enter the linear regime set by
\begin{eqnarray}
  \label{length}
  FL > \Delta_1.
\end{eqnarray}  
In this case the standing wave far away on the left from the tunnelling
region is best approximated as the solution of an effective linear Dirac equation. Full treatment of the Dirac equation is presented in Appendix \ref{sec:Dirac}, here present some necessary steps to obtain the asymptotic standing wave solution.  Assuming a linear, gapped 2D Dirac equation subject to external field, 
\begin{eqnarray}
  \label{HD}
  \left[v({\bm p}\cdot{\bm \sigma})+\Delta\sigma_z-Fx\right]\psi=\varepsilon\psi
\end{eqnarray}
the two-component wave function takes the same structure as \eqref{wavefunc},  yet the $x$-component of the probability current is now
\begin{eqnarray}
  \label{jx}
j_x=v\left[a(x)^*b(x)+b(x)^*a(x)\right].
\end{eqnarray}  
Considering $\varepsilon=0$, the semiclassical $x$-momentum is
\begin{eqnarray}
  p(x)&=&\pm\frac{1}{v}\sqrt{(F x)^2-\Delta_k^2}\nonumber\\
\Delta_k&=&\sqrt{\Delta^2+v^2k_y^2}.
\end{eqnarray}
There are two turning points where $p(x)=0$,
$x_{\pm}=\pm \Delta_k/F$. Here we are only interested in the standing wave at $F|x| \gg \Delta,vk_y$, which takes the asymptotic form (see Appendix~\ref{sec:Dirac} for details)
\begin{eqnarray}
  \label{sw}
  \varphi&=& \left(
  \begin{array}{l}
   \cos (S_- + \phi)    \\
    -i\sin(S_- + \phi)
  \end{array}
  \right)
\nonumber\\
S_-&=&\int_{x_-}^{x}p_{x^\prime} dx^\prime.
\end{eqnarray}  
The standing wave, $\varphi$, is our asymptotic solution for Bblg subject to $FL > \Delta_1$.
This wave function is normalised per one particle in unit area,
which is the correct normalisation of the Dirac sea.
When approaching the tunnelling region this wave is transformed to (\ref{sw1}).
The normalisation of  (\ref{sw1}) must correspond to (\ref{sw}).
To match (\ref{sw}) with (\ref{sw1}) consider a running component of the standing wave, say
the component running from the left to the right.
According to (\ref{jx}) the running component of (\ref{sw}) carries the current $j_x=v/2$.
On the other hand, according to (\ref{jx1}) the running component (\ref{sw1}) carries the current
$j_x=\frac{1}{2m}$. 
Hence, all the  wave functions in the bilayer section must be multiplied by $\sqrt{mv})$ to bring
it to the correspondence with (\ref{sw}). Hence the running escaping wave (\ref{rw1}) carries the current
$j_x=(mv) |T|^2/(2m)=v|T|^2/2$.

\subsubsection{Pair production rate in the linear regime}

\begin{figure}[t!]
  \includegraphics[width=0.45\textwidth]{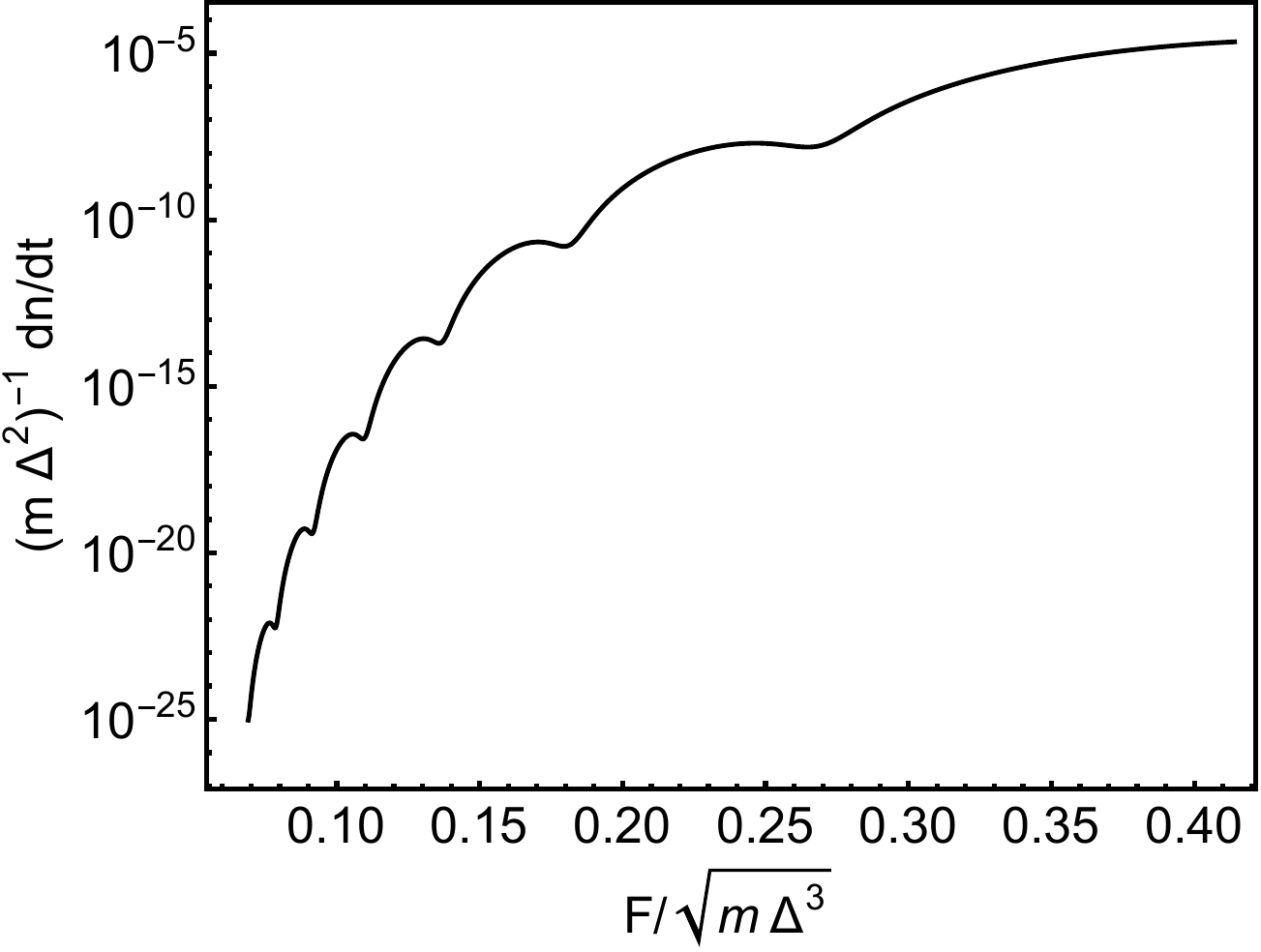}
  \caption{Pair production rate per unit time and area, rescaled by $(m\Delta^2)^{-1}$, plotted against the dimensionless field $F/\sqrt{m\Delta^3}$. This rescaling renders $(m\Delta^2)^{-1} \, dn/dt$ both dimensionless and universal, in the sense that the pair production depends only on the single combination $F/\sqrt{m\Delta^3}$.
}
\label{fig:dn}
\end{figure}

Again, the wave function (\ref{sw}) asymptotically, $x \to -\infty$  is normalised as
$|a|^2+|b|^2=1$. 
Therefore the $x$-density of states is
\begin{eqnarray}
  \label{rhox}
  \rho_x=\frac{dp_x}{\pi}=\frac{d\varepsilon}{\pi v}= \frac{Fdx}{\pi v}.
\end{eqnarray}
Note that we divide by $\pi$ instead of $2\pi$ because this is the standing wave.
We also account that $d p_x=d(vp_x)/v=d\epsilon/v=Fdx/v$, here $dx$ is the length
that corresponds to the energy change $d\varepsilon$. 
The $y$-density of states is usual
\begin{eqnarray}
  \label{ry}
  \rho_y=\frac{dy dk_y}{2\pi}
\end{eqnarray}
Together with (\ref{rhox}) this gives the following
number of the decay events per unit time per unit area, with area $A$ and number density $n=N/A$, 
\begin{align}
  \label{ndotbl}
  &\frac{dn}{dt}=\frac{F}{2\pi}
  \int_{-\infty}^{\infty}|T(k_y)|^2\frac{dk_y}{2\pi}\\
   \notag &=
   \frac{(m\Delta^2)e^{-\sqrt{2}S_0}}{2\pi(\pi\beta_1)^{1/2}}\left(\frac{F^2}{m\Delta^3}\right)^{3/4} \left[1-\frac{\cos(\sqrt{2}S_0-\pi/8)}{2^{1/4}}\right].
\end{align}
This has to be multiplied by the spin-valley degeneracy, $g_{s\tau}=4$. A representative plot is provided in Fig. \ref{fig:dn}. 

We stress that the form of the exponent in Eq.~\eqref{ndotbl} is straightforward to compute, but that the prefactor is non-trivial and requires careful wavefunction matching—as we have done—as well as correct normalisation of the standing-wave asymptotics. Having established the properly normalised pair production rate, one can compute the current as $I = L W e\, dn/dt$, for a device of width $W$, length $L$, and with $e$ the electric charge.

\subsubsection{Normalisation and tunnelling rate in the quadratic regime (short  device)}
Consider now the {\it short device} such that 
\begin{eqnarray}
  \label{length1}
 \Delta \ll FL \ll \Delta_1.
\end{eqnarray}  
According to Eq.~(\ref{sw1}) the wave function at $x\to -\infty$ is
\begin{eqnarray}
  \label{sw3}
  \varphi&=& \frac{1}{\sqrt{p_{-\infty}}}\cos (S+\phi)
    \left(
  \begin{array}{c}
   1    \\
    -1
  \end{array}
  \right)
\end{eqnarray}  
  where $p_{-\infty}$ is momentum at $x\to -\infty$.
  Hence, to get the necessary normalization one particle per unit length
  one need to multiply the wave fanction by $\sqrt{p_{-\infty}}$.
  This results in the following escaping current per one quantum state
  $j_x=\frac{|T|^2}{2m}p_{-\infty}$. To get the  total current
  one needs to multiply $j_x$ by density of states given by Eq. (\ref{rhox}), with the velocity taken as $v= v_{-\infty}$. Since $FL \gg \Delta$ the dispersion at
  $x=\pm\infty$ is quadratic, $v_{-\infty}=p_{-\infty}/m$. All-in-all
  this gives the same result for the number of the decay events per unit time per unit area as that for the long device case, Eq.~(\ref{ndotbl}).

\section{Discussion}\label{sec:discussion}

\subsection{Oscillatory tunnelling and localised modes}

We reiterate a key feature noted in the earlier work of Nandkishore and Levitov~\cite{LevitovPNAS2011}, and later argued to be generic in systems with higher-than-quadratic dispersion~\cite{Johri2013}: the tunnelling current exhibits an oscillatory dependence on both the applied electric field and the gap parameter, reflecting interference effects intrinsic to interband tunnelling in such systems.

Building on this, our analytic treatment reveals an additional, notable feature: the presence of an exponentially decaying (evanescent) component in the classically allowed region, schematically denoted as $B|e^{-S}\rangle$ in Eq.~\eqref{lcr}. These decaying solutions are essential to satisfy consistent matching conditions. Such components have been rigorously analysed in earlier semiclassical work, notably by Voros~\cite{Voros1983}. 

For special values of transverse momentum $k_y^0$, the transmission vanishes, $T(k_y^0) = 0$, yet the coefficient $B$ remains nonzero (see Eq.~\eqref{leq}). This implies the existence of a purely localised mode at fixed $k_y^0$: a solution with zero transmission but finite amplitude, confined near the classical turning points. This behaviour is reminiscent of total internal reflection in optics, where an incident wave is fully reflected, yet an evanescent tail penetrates into the adjoining medium—not classically forbidden, but still unable to support propagating solutions for the given transverse momentum.

Finally, we speculate that these decaying solutions may offer a fruitful analogy to non-Hermitian physics, where complex momenta and non-unitary dynamics similarly lead to boundary-localised modes~\cite{Yao2018,Bergholtz2021}.

\subsection{Summary}

We considered biased bilayer graphene—i.e., with a finite band gap—subject to an in-plane electric field, which induces tunnelling across the gap. We presented a semiclassical method based on analytic continuation, yielding fully analytic expressions for the transmission probability. This approach avoids reliance on exact solutions near turning points—such as Airy functions in the quadratic case—which are unlikely to exist for quartic dispersions. It also reveals a peculiar structure in the wavefunctions, including essential evanescent components. Finally, using this solution, we obtained the absolute normalisation of the pair-production rate and, with it, the correct expression for the tunnelling current.


%

\appendix

\section{2D GAPPED DIRAC}\label{sec:Dirac}
\subsection{$x>x_+$}
Here we consider the escaping wave, $p_A(x)>0$, and solve (at $\varepsilon=0, k_y=0$)
\begin{align}
    (p_A(x)\sigma_x + \Delta \sigma_z- F x\sigma_0) \varphi_R = 0.
\end{align}
The full wavefunction $\Psi_R=\varphi_R e^{iS_+}$ is given by
\begin{align}
\label{PsiR}
   \Psi_R&=\sqrt{\frac{J}{2v}}\sqrt{\frac{F}{p_A(x)}}\begin{pmatrix} (x-x_-)^\frac{1}{2} \\ (x-x_+)^\frac{1}{2}\end{pmatrix} e^{iS_+}.
\end{align}
Normalisation  is such the the current along $x$ is $J$. 

\subsection{$x_-<x<x_+$}
Starting at $x\sim x_+$, we perform acl and cl continuation via $x-x_+=\rho e^{\pm i\pi}$, respectively. Under this,
\begin{align}
   \text{acl}: \ \ &(x-x_+)^\frac{1}{2} \to i (x_+-x)^\frac{1}{2},\\
    \text{cl}: \ \ &(x-x_+)^\frac{1}{2} \to -i (x_+-x)^\frac{1}{2}
\end{align}
and 
\begin{align}
   \text{acl}: \ \ &i S_+ \to -S_+,\\
    \text{cl}: \ \ &i S_+ \to S_+.
\end{align}
Running from right to left in $x$, we only keep solutions that are exponentially growing in the forbidden region, i.e. only the cl continuation. Finally, it is more convenient to work with respect to the left turning point, $x_-$, and hence we use the relation $S_+=S_0 - S_-$. All-in-all, the continued wavefunction is
\begin{align}
   \Psi_F(x)&=\sqrt{\frac{J}{2v}}\sqrt{\frac{F}{-ip_F(x)}}\begin{pmatrix} (x-x_-)^\frac{1}{2} \\ -i(x_+-x)^\frac{1}{2}\end{pmatrix} e^{S_0-S_-}.
\end{align}

\subsection{$x<x_-$}
From $\Psi_F(x)$ we analytically continue in $x-x_-\to(x_--x)e^{\pm i\pi}$, with sign corresponding to acl and cl, respectively. Since we desire a standing wave for $x<x_-$, we will need to keep both paths in the complex plain. Explicitly, these paths give
\begin{align}
   \text{acl}: \ \ &-S_- \to -iS_-,\\
    \text{cl}: \ \ &-S_- \to iS_-.
\end{align}
which follows from
\begin{align}
   \text{acl}: \ \ &(x-x_-)^\frac{1}{2} \to i (x_--x)^\frac{1}{2},\\
    \text{cl}: \ \ &(x-x_-)^\frac{1}{2} \to -i (x_--x)^\frac{1}{2}
\end{align}
and, importantly, we must keep track of the global coefficient, i.e. containing $p_F(x)$, which continues to 
\begin{align}
   \text{acl}: \ \ &(-i p_F(x))^\frac{-1}{2} \to ( p_A(x))^\frac{-1}{2},\\
    \text{cl}: \ \ &(-i p_F(x))^\frac{-1}{2} \to -i( p_A(x))^\frac{1}{2}
\end{align}

The full wavefunction is given by $\Psi_L=\Psi^\text{acl}_L+\Psi^\text{cl}_L$. For ease, we multiply by global phase factor exp$(i\pi/4)$
\begin{align}
   \Psi^\text{acl}_L&=i\sqrt{\frac{J}{2v}}\sqrt{\frac{F}{p_A(x)}}\begin{pmatrix} (x_--x)^\frac{1}{2} \\ -(x_+-x)^\frac{1}{2}\end{pmatrix} e^{S_0-iS_-+i\frac{\pi}{4}},\\
   \Psi^\text{cl}_L&=i\sqrt{\frac{J}{2v}}\sqrt{\frac{F}{p_A(x)}}\begin{pmatrix} -(x_- -x)^\frac{1}{2} \\ -(x_+-x)^\frac{1}{2}\end{pmatrix} e^{S_0+iS_--i\frac{\pi}{4}},
\end{align}
we therefore arrive at
\begin{align}
\label{PsiL}
   \Psi_L&=-\sqrt{\frac{2J}{v}}e^{S_0}\sqrt{\frac{F}{p_A(x)}}\begin{pmatrix} (x_--x)^\frac{1}{2} \cos(S_-+\pi/4) \\ (x_+-x)^\frac{1}{2}(-i)\sin(S_-+\pi/4)\end{pmatrix}.
\end{align}

\subsection{Current}
Deep in $x\ll x_-$, the standing wave is normalised to unity (with $A=$area), and hence we can find the current $J$,
\begin{align}
   \frac{1}{A}||\Psi_L||^2&\to \frac{2J}{v}e^{2S_0} = 1
\end{align}
and hence
\begin{align}
\label{JDirac}
   J = \frac{1}{2} v e^{-2S_0}.
\end{align}

\begin{figure}[t!]
\includegraphics[width=0.4\textwidth]{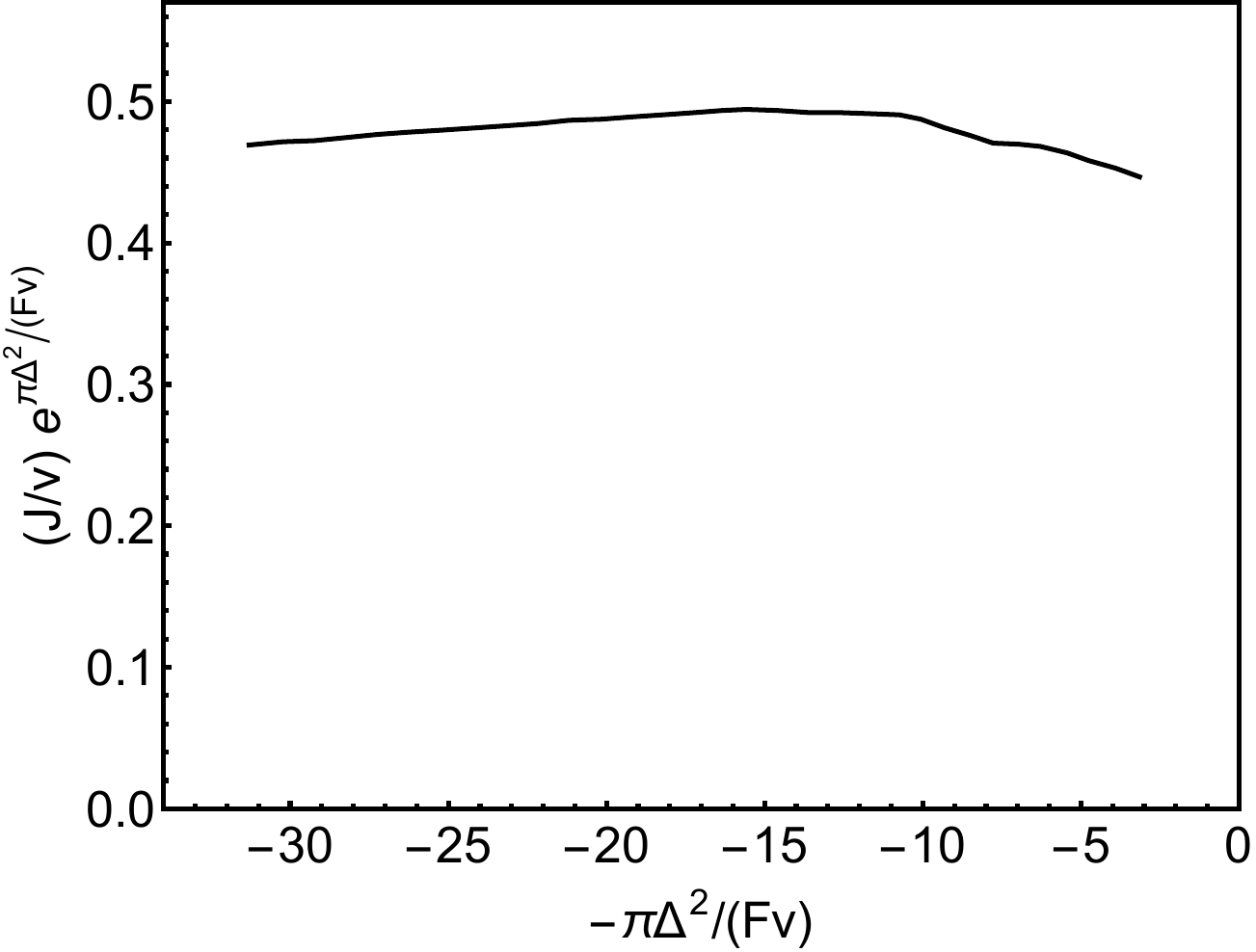}
\caption{Numerical evaluation of the tunnelling current $J$. Numerical differential equation is solved using the simplest Euler method and the procedure is stable within the range
$2.3\times 10^{-14}< e^{-\pi\Delta_k^2/vF}< 4\times10^{-2}$.}
\label{coef}
\end{figure}

One can verify the current \eqref{JDirac} by direct numerical methods. Here we solve the Dirac Equation, evolving under $F x =i F \partial_{p_x}$.
Starting from the asymptotic (\ref{PsiR}), we run the solution from large positive $x$  towards large negative $x$
where the numerical solution  coincides with form of (\ref{PsiL}), but with coefficient left unknown, i.e.
\begin{align}
\label{PsiLdash}
   \Psi_L'&=C\sqrt{\frac{F}{p_A(x)}}\begin{pmatrix} (x_--x)^\frac{1}{2} \cos(S_-+\pi/4) \\ (x_+-x)^\frac{1}{2}(-i)\sin(S_-+\pi/4)\end{pmatrix}.
\end{align}
with $C$ the unknown coefficient. Matching $C$ to that found by numerically evolving \eqref{PsiR} to large negative $x$, we find that, within numerical accuracy 
\begin{align}
    C\approx \sqrt{\frac{2J}{v}}e^{S_0}.
\end{align}
The plot of $J/v \exp(\pi\Delta_k^2/vF)$ vs $-\pi\Delta_k^2/vF$ is presented
in Fig.~\ref{coef}.  

\subsection{Pair-production rate}
The wave function (\ref{PsiL}) asymptotically, $x \to -\infty$  is normalised as
$|a|^2+|b|^2=1$. 
Therefore the $x$-density of states is
\begin{eqnarray}
  \label{rx}
  \rho_x=\frac{dp_x}{\pi}=\frac{d\epsilon}{\pi v}= \frac{Fdx}{\pi v}
\end{eqnarray}
Note that we divide by $\pi$ instead of $2\pi$ because this is the standing wave.
We also account that $d p_x=d(vp_x)/v=d\epsilon/v=Fdx/v$, here $dx$ is the length
that corresponds to the energy change $d\epsilon$. 
The $y$-density of states is usual
\begin{eqnarray}
  \label{ry}
  \rho_y=\frac{dy dk_y}{2\pi}
\end{eqnarray}
Hence, the number of the decay events per unit time is
\begin{align}
  \label{ndot1}
        \frac{dN}{dt}= \frac{F dx}{\pi v} e^{-\pi\Delta^2/vF}\int_{-\infty}^{+\infty}
        \frac{1}{2}ve^{-\pi k_y^2 v/F}\frac{dy dk_y}{2\pi}
\end{align}
This gives the following number of the decay events per unit time per unit area
\begin{align}
  \label{ndot2}
  \frac{dN}{dt dx dy}=\frac{1}{(2\pi)^2}\left(\frac{vF}{\Delta^2}\right)^{3/2}
\Delta\left(\frac{\Delta}{v}\right)^2\exp\left(-\frac{\pi \Delta^2}{vF}\right)
\end{align}
This has to be multiplied by the spin-valley degeneracy. In the main text we use notation $dn/dt = dN/(dt dx dy)$.

\clearpage

\end{document}